\documentclass[%
 aip,
 amsmath,amssymb,
 reprint,%
]{revtex4-1}
\usepackage{color}
\usepackage{graphicx}
\usepackage{dcolumn}
\usepackage{bm}

\usepackage[utf8]{inputenc}
\usepackage[T1]{fontenc}
\usepackage{mathptmx}
\usepackage{etoolbox}
\UseRawInputEncoding
\makeatletter
\def\@email#1#2{%
 \endgroup
 \patchcmd{\titleblock@produce}
  {\frontmatter@RRAPformat}
  {\frontmatter@RRAPformat{\produce@RRAP{*#1\href{mailto:#2}{#2}}}\frontmatter@RRAPformat}
  {}{}
}%
\makeatother
\begin{document}

\preprint{AIP/123-QED}

\title{Continuous-frequency weak electric field measurement with Rydberg atoms}
\author{Jinlian Hu}
 
\author{Huaqiang Li}

\author{Rong Song}
 
\author{Jingxu Bai}
 \affiliation{State Key Laboratory of Quantum Optics and Quantum Optics Devices, Institute of Laser Spectroscopy, Shanxi University, Taiyuan 030006, P. R. China
}%

\author{Yuechun Jiao}
\email{ycjiao@sxu.edu.cn}

\author{Jianming Zhao}
 \email{zhaojm@sxu.edu.cn}

\author{Suotang Jia}
 \affiliation{State Key Laboratory of Quantum Optics and Quantum Optics Devices, Institute of Laser Spectroscopy, Shanxi University, Taiyuan 030006, P. R. China
 }%
 \affiliation{Collaborative Innovation Center of Extreme Optics, Shanxi University, Taiyuan 030006, China}

\date{\today}

\begin{abstract}
   We demonstrate a continuous frequency electric field measurement based on the far off-resonant  AC stark effect in a Rydberg atomic vapor cell. In this configuration, a strong far off-resonant field, denoted as a local oscillator (LO) field, acts as a gain shifting the Rydberg level to a high sensitivity region. An incident weak signal field with a few hundreds of kHz difference from the LO field is mixed with the  LO field in Rydberg system to generate an intermediate frequency (IF) signal, which is read out by the Rydberg electromagnetically induced transparency (Rydberg-EIT) spectroscopy. Not like resonant EIT-AT spectra, we realize the electric field  measurement of the signal frequency from 2 GHz to 5 GHz using a single Rydberg state. A minimum detectable filed strength is down to 2.31~$\mu$V/cm and a linear dynamic range is over 65~dB. The minimum detectable filed is comparable with a resonant microwave-dressed Rydberg heterodyne receiver using the same system, which is 1.45~$\mu$V/cm. We also show the system has an inherent polarization selectivity feature. Our method can provide a high sensitivity of electric field measurement and be extended to arbitrary frequency measurements.  
\end{abstract}


\begin{quotation}
\end{quotation}

\maketitle

\section{Introduction}
    An atom-based sensing has unique advantages in the measurement of weak signals with   high sensitivity, calibration-free, and intrinsic accuracy. Seminal advances include atomic clocks ~\cite{1} and magnetometers ~\cite{2}. A quantum sensor for the radio frequency (RF) field is also expected to follow because of its widespread application. 
	
	A  significant progress has been made in the atom-based measurement of microwave electric fields using Rydberg atoms due to their large dc polarizabilities and microwave-transition dipole moments ~\cite{3}. An optical Rydberg electromagnetically induced transparency (Rydberg-EIT) spectroscopy and $\mbox{Aulter-Townes}$(AT) splitting have been employed to measure the RF field over a wide frequency range from 100~MHz ~\cite{4,5} to over 1 THz ~\cite{6}, including the measurement of the strength of microwave field and polarizations ~\cite{7,8,9}, the sub-wavelength microwave imaging ~\cite{10,11} and the first demonstration of angle-of-arrival measurements ~\cite{12}. The concept of the wireless communication recently based on Rydberg atoms also has been demonstrated, including an amplitude modulation (AM) ~\cite{13,14,15}, a frequency modulation (FM) ~\cite{16}, and a phase modulation (PM) ~\cite{17,18}. Recently, the state-of-the-art Rydberg atomic sensor is achieved using a heterodyne receiver architecture with a sensitivity of 55 nVcm$^{-1}$Hz$^{-1/2}$ ~\cite{19,20}. However, the above mentioned works have relied on resonant or near-resonant transitions between Rydberg states, which restricts any measurement of microwave fields to a discrete set of atomic transition frequencies from $\thicksim$ GHz to $\thicksim$ THz with a bandwidth less than 10MHz. Using all of the available transitions requires a laser system with broad tunability or multiple laser system realizing frequency stabilization to atomic transitions, which is complex and expensive. Distinct from the resonant EIT-AT method, a sensor based on the AC Stark shift can achieve a continuous-frequency measurement. The proof-of-principle experiment indicates that the Rydberg sensor can achieve the continuous-frequency measurement for sufficiently strong electric field from 200~V/m to $\textgreater$ 1~kV/m~\cite{21,22}. In order to achieve the continuous-frequency weak electric field measurement, an off-resonance heterodyne technique is utilized to greatly boost the sensitivity at arbitrary frequencies in the waveguide-coupled Rydberg system~\cite{23}, while it is not likely to achieve absolute electric field accuracy outperforms previous free-space Rydberg electric field measurements.
	
	In this work, we present an AC-Stark based weak RF electric field measurement technique with the RF frequency from 2~GHz to 5~GHz using one Rydberg level in a vapor cell. The basic idea is that a strong far off-resonant field is introduced as a local oscillator (LO) field to AC Stark shift the Rydberg level to the high sensitivity point. Specifically, we choose $|60D_{5/2}\rangle$ state that exhibits $m_{j}$-dependent AC stark shifts and splittings in the presence of a strong LO field. We lock the laser frequency to a big slope point of the $|m_j=1/2\rangle$ Stark level because of its larger polarizability than that of $m_j = 5/2, 3/2$. The incident weak signal field, mixed with the LO field in Rydberg system, is read out by Rydberg-EIT. The read-out signal is linear dependence on the strength of the signal field with a dynamic range over 65dB, and the minimum detectable field strength is down to 2.31~$\mu$V/cm, which is comparable with a resonant microwave-dressed Rydberg heterodyne receiver, achieving 1.45~$\mu$V/cm with the same experimental system. We also show the polarization detection of the RF field, corresponding isolation ratio for two orthogonal signal fields being up to 32~dB. Our method, in principle,  provides the RF electric field measurement at arbitrary frequencies with a high sensitivity.

\section{Experimental setup}
    \begin{figure}[htbp]
		\vspace{-1ex}
		\centering
		\includegraphics[width=0.5\textwidth]{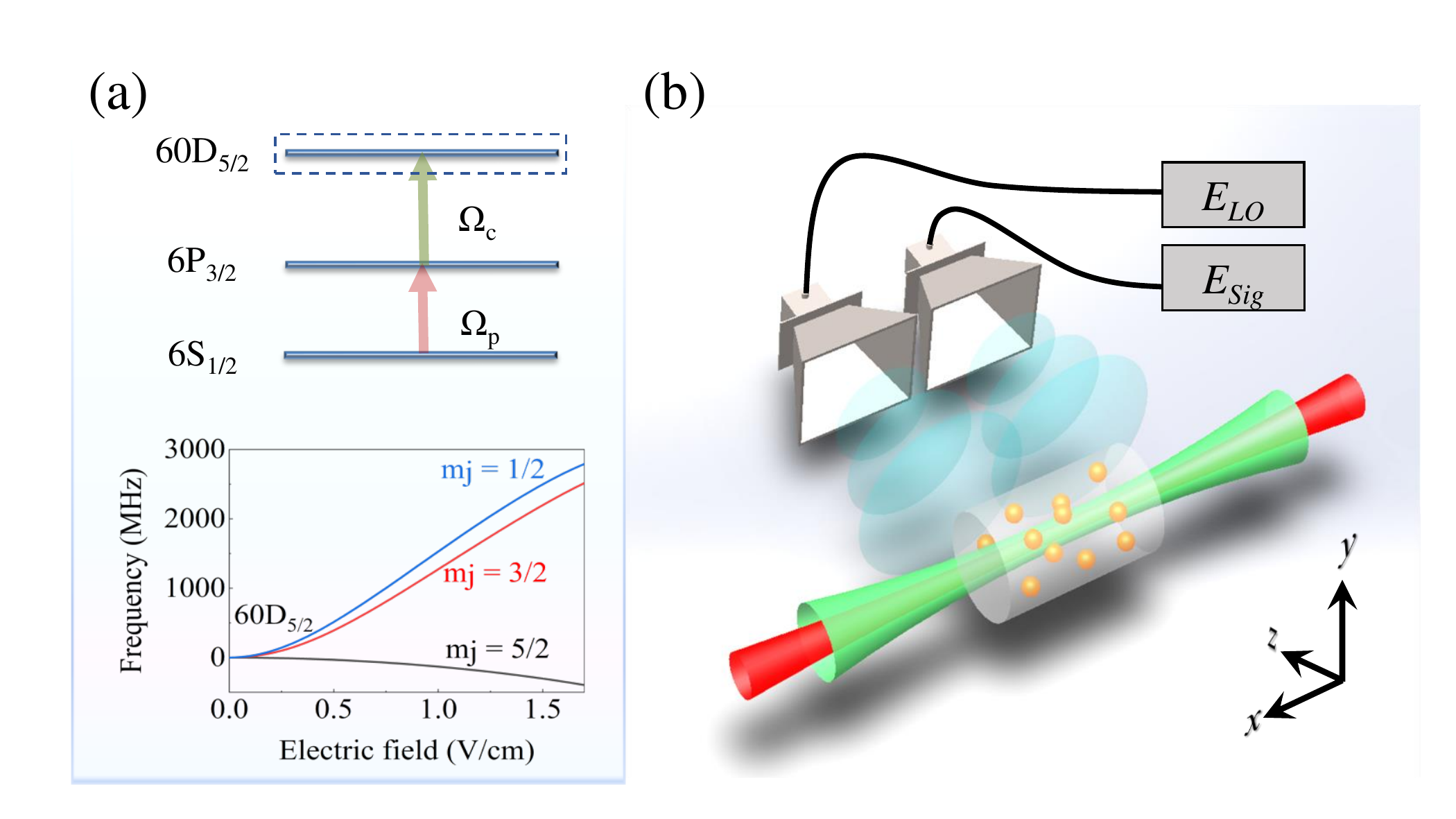}
		\vspace{-1ex}
		\caption{(a) Top: Energy-level diagram of the three-level Rydberg-EIT of a cesium atom. A probe laser, $\Omega_p$ ($\lambda_p$ = 852~nm), is resonant with the lower transition $|6S_{1/2}(F = 4)\rangle\to |6P_{3/2}(F^\prime = 5)\rangle$, and a coupling laser $\Omega_c$ ($\lambda_c$ = 510~nm) couples the  transition of $|6P_{3/2}(F^\prime = 5)\rangle\to |60D_{5/2}\rangle$. Below: Stark spectra of the Rydberg state  $|60D_{5/2}\rangle$. With increasing the strength of field, the energy level shifts and splits into three  Stark levels, $|m_{j} = 5/2\rangle$ (black line), $|m_{j} = 3/2\rangle$ (red line), $|m_{j} = 1/2\rangle$ (blue line).  (b) Sketch of the experimental setup. A coupling laser and a probe laser counter-propagate through a cylindrical room-temperature cesium cell. The transmission of the probe laser is detected using a photodiode (PD). A reference beam (not shown in here), separated from the probe laser beam, is subtracted with the transmission probe laser to eliminate any probe laser power drift. A LO field $E_{LO}$ and a weak signal field $E_{Sig}$ are emitted from two identical horn antennas.}
		\label{Fig.1}
	\end{figure}
	
	The experimental setup and the relevant $^{133}$Cs Rydberg-EIT three-level diagram is illustrated in Fig.~\ref{Fig.1}. A Rydberg coupling laser ($\lambda_c$ = 510~nm) and a probe laser ($\lambda_p$ = 852~nm) counter-propagate through a cylindrical room-temperature cesium cell with 50~mm long and 20~mm diameter, where the probe beam with a $1/e^2$ waist radius of 500~$\mu$m and a power of 200~$\mu$W, is locked to the transition of $|6S_{1/2}(F = 4)\rangle\to |6P_{3/2}(F^\prime = 5)\rangle$ using the modulation transfer spectroscopy~\cite{24}, and the coupling laser with a $1/e^2$ waist radius of 550~$\mu$m and a power of 18.5~mW further excites the atoms to the Rydberg state $|60D_{5/2}\rangle$ thus establishing the EIT to enhance the probe transmission. The EIT signal is detected by measuring the transmission of the probe laser using a photodiode (PD). A reference beam, separated from the probe laser beam, is subtracted with the transmitted EIT signal to eliminate any probe laser power drift. The probe and coupling lasers keep co-linear polarization along y-axis.

	Two radio frequency fields with the co-linear polarization along y-axis, denoted as a local oscillator (LO) field  $E_{LO}$  and a weak signal fields $E_{Sig}$, incident to the Rydberg system simultaneously. The LO field operates at frequency of 2.18 GHz (SRS SG386), which is far off-resonant with the transition of $|60D_{5/2}\rangle$ to adjacent $|61P_{3/2}\rangle$ Rydberg state, and emitted from a horn antenna (A-info LB-20180-SF). Under the strong LO field, the $|60D_{5/2}\rangle$ Rydberg state exhibits $m_{j}$-dependent ($m_{j}$ = 1/2, 3/2, 5/2) AC Stark shifts and splittings, shown as the below of the Fig.~\ref{Fig.1} (a). The slope of $m_{j}$ = 1/2 Stark level is much larger than that of zero field, which means the $m_{j}$ = 1/2 Stark level is more sensitivity to the electric field. The weak signal field with 75~kHz frequency difference of LO field is generated with the other signal generator and emitted with the other identical horn antenna.
	
\section{Results and discussion}
	When the LO field ($E_{LO}$) and the signal field ($E_{Sig}$) are incident on the vapor cell, the response of the Rydberg atom to both fields is described in detail in the Ref.~\cite{23}. Here, we give the simplified results. 
	 
	For the near resonance field of the Aulter-Townes regime, the total microwave electric field ($E_{tot}$) that the atoms feel can be expressed as 
    \begin{equation}\label{Eq.1}
		E_{tot} \approx E_{LO}/2+E_{Sig}/2cos(\Delta \omega +\Delta\phi),
	\end{equation}
	and the Aulter-Townes splitting is given by
	\begin{equation}\label{Eq.2}
		\Delta{f}_{AT}=2\pi\frac{\mu}{\hbar}E_{tot},
	\end{equation}
	where $E_{LO}$($E_{Sig}$), $\omega_{LO}$($\omega_{Sig}$), $\phi_{LO}$($\phi_{Sig}$) represent the amplitude, angular frequency and phase of the local oscillation field (signal field), respectively. $\Delta \omega$ is the difference between the angular frequency of the local field and the signal field, $\Delta\omega=\omega_{LO}-\omega_{Sig}\ll\overline{\omega} = (\omega_{LO}+\omega_{Sig})/2$. $\Delta\phi$ is the difference between the phase of the local field and the signal field, $\Delta\phi=\phi_{LO}-\phi_{Sig}$. $\Delta{f}_{AT}$ is the microwave field dressed  Aulter-Townes splitting, $\hbar$ is reduced Planck constant, $\mu$ is the transition matrix element between two Rydberg states. 

    For the far off-resonant of Stark shift regime, the total field the Rydberg atoms experienced becomes 
    \begin{equation}\label{Eq.3}
		E^2_{tot} \approx E^2_{Sig}/2+E^2_{LO}/2+E_{Sig}E_{LO}cos(\Delta \omega+\Delta\phi),
	\end{equation} 
    and the Stark shift of the Rydberg state depends on the atomic polarizability, which is written as,
    \begin{equation}\label{Eq.4}
		\Delta{f}_{stark}=-\frac{1}{2}\alpha\langle{E^2_{tot} }\rangle, 
	\end{equation}
       where $\Delta{f}_{stark}$ is the Stark shift, $\alpha$ is the atomic polarizability of AC Stark effect, $\langle{E^2_{tot}}\rangle$ is the average value of the square of the electric field.

\begin{figure}[htbp]
		\vspace{-1ex}
		\centering
		\includegraphics[width=0.5\textwidth]{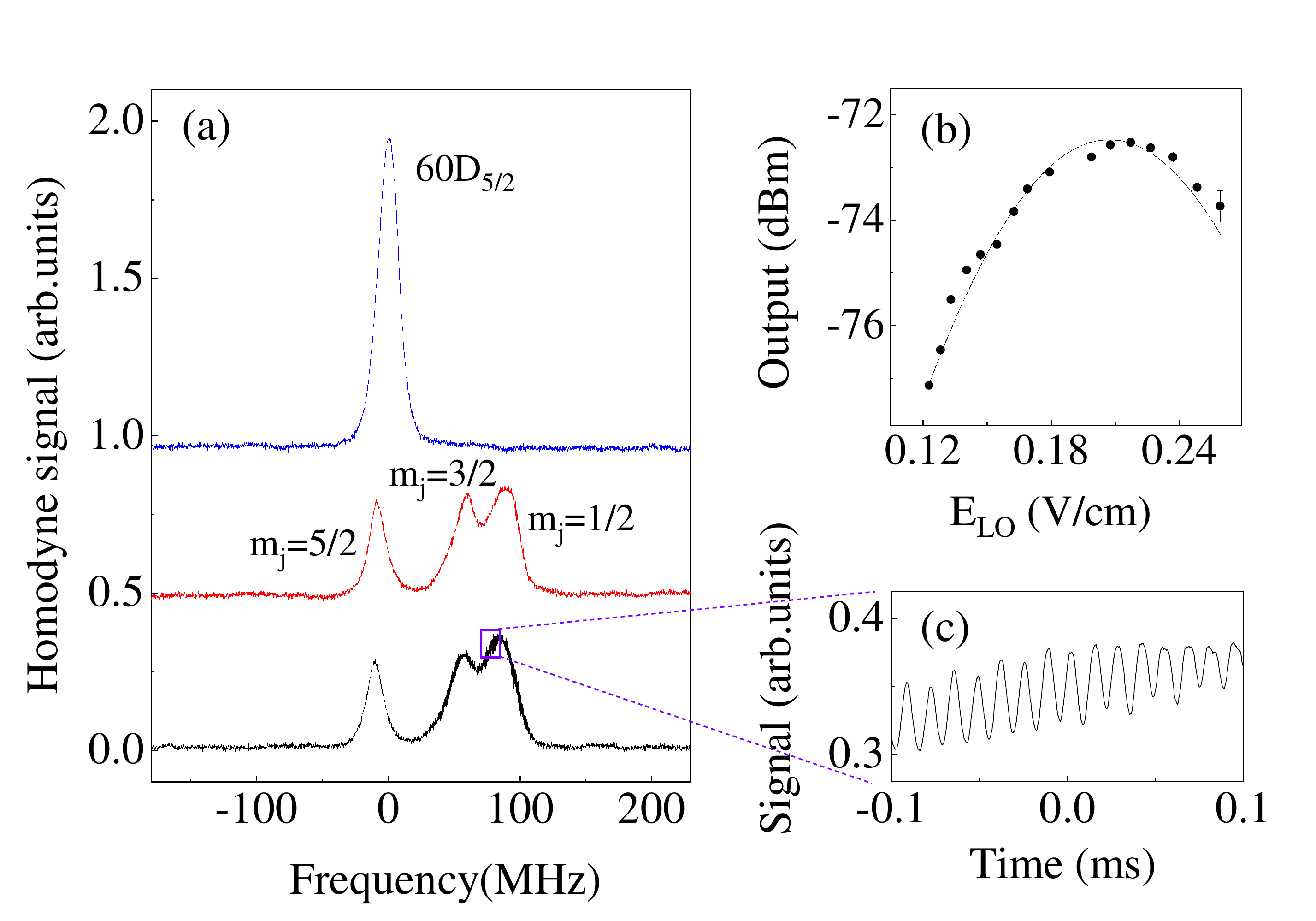}
		\vspace{-1ex}
		\caption{(a)  Rydberg-EIT spectra for $|60D_{5/2}\rangle$ state without a RF field (the top curve), with only a LO field (the middle curve), and with both the strong LO field and a weak signal simultaneously (the bottom curve). The main peak at near 0 detuning is formed by the cascade three-level atom of $|6S_{1/2}(F = 4)\rangle\to |6P_{3/2}(F^\prime = 5)\rangle\to |60D_{5/2}\rangle$, marked with a dashed line. In the presence of the LO field, EIT spectrum shows energy shifts and splits into three peaks due to AC Stark effect, denoted as $m_j$ = 1/2, 3/2, 5/2. When both LO field and signal field are incident, the EIT spectrum further shows amplitude-modulation at frequency $\Delta \omega/2\pi$ on the spectrum line. A maximal amplitude-modulated response appears at $m_j $ = 1/2 Stark level due to its larger polarizability. (b)The response of system to the LO field with fixed signal field. (c) The zoom part for the spectrum of $m_j$ = 1/2 Stark level in time domain with modulation frequency $\Delta \omega/2\pi$ = 75~kHz.}
		\label{Fig.2}
	\end{figure}

    In both cases, the Rydberg-EIT signal varies at frequency of $\Delta \omega/2\pi$ and the related amplitude is linear proportional to the signal field $E_{Sig}$. Especially, under the far off-resonant red case, the LO field $E_{LO}$ acts as a gain to amplify the signal field, shown with the last term of Eq.~3. In Fig.~\ref{Fig.2} (a), we demonstrate Rydberg-EIT spectra for $|60D_{5/2}\rangle$ state without a RF field (the top curve), with only the LO field  (the middle curve), and with both the strong LO field and the weak signal (the bottom curve). The LO field is  $E_{LO}$ = 0.18~V/cm with the frequency of $f_{LO}$ = 2.18~GHz, whereas the signal field is  $E_{Sig}$ = 7.23~mV/cm with the frequency of $f_{Sig}$ = 2.180075~GHz, having a fixed frequency difference $\delta$ = 75~kHz. It is clearly shown that in the presence of the local field, the EIT spectrum shows energy shifts and split into three peaks, denoted as $m_j$ = 1/2, 3/2, 5/2, as shown with the middle curve. When both the LO field and the weak signal interact with atoms, we can observe the Rydberg-EIT line broadening, as shown with the bottom curve of Fig.~\ref{Fig.2} (a). The $m_j$ = 1/2 Stark spectrum exhibits a maximal  broadening comparing to the $m_j$ = 5/2, 3/2 Stark lines due to the larger polarizability of $m_j$ = 1/2 line, see the enlargement of the square marked region in Fig.~\ref{Fig.2} (c). This EIT line broadening is attributed to the beat signal between the LO field and the signal field~\cite{25}. From Eq.~\ref{Eq.3}, the weak signal, $E_{Sig}$, can be obtained by analyzing the modulated EIT spectrum with a spectrum analyzer (ROHDE $\&$ SCHWARZ FSVA13).     

    In order to measure the weak signal field, the center of the $m_j$=1/2 Stark line is chosen as the 510~nm laser frequency operating point forming Rydberg EIT. In Fig.~\ref{Fig.2} (b), we present the sensitivity test of our Rydberg sensor, where we measure the output signal of the spectrum analyzer for different  power of the LO field and fixed  signal field $E_{Sig}$ = 0.73~mV/cm. We can see the LO field has an optimal operating point, where the system is the most sensitivity for the signal field. In the following, we fix the strength of the LO field operating at the optimal point $E_{LO}$ = 0.21~V/cm.
		
\begin{figure}[htbp]
		\centering
		\vspace{-1ex}
		\includegraphics[width=0.5\textwidth]{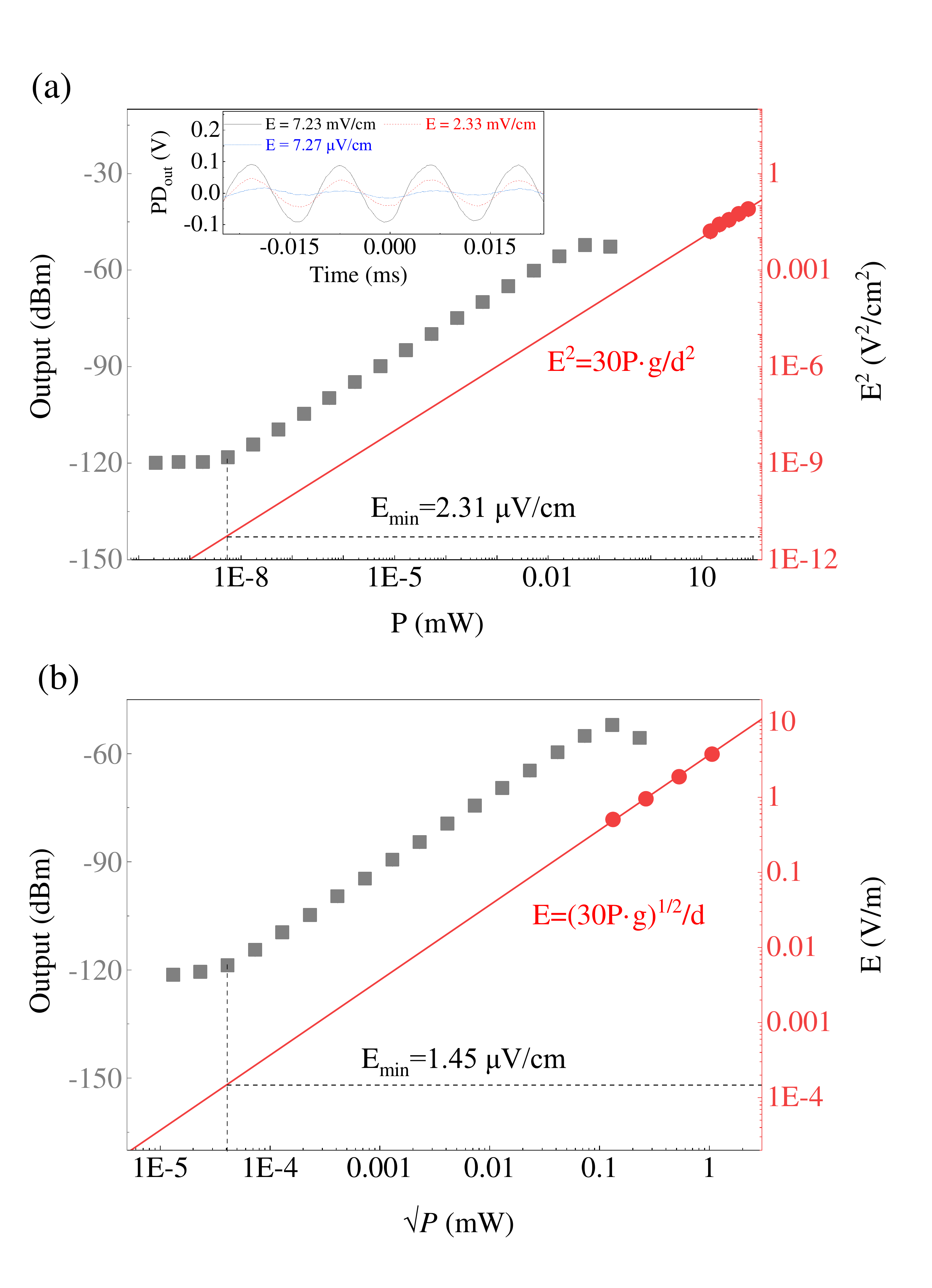}
		\vspace{-1ex}
		\caption{(a) Measurement of the weak signal field using the AC Stark technique with $f_{LO}$ = 2.18~GHz, and $\delta$ = 75~kHz. The data is taken with the far off-resonant AC stark technique (gray squares), and with the AC Stark shift in strong regime (red dots). The red solid line displays the electric field calculated by Eq.~\ref{Eq.5}. The dash line shows the minimum measured electric field of $E_{min}$ = 2.31~$\mu$V/cm. Inset: the probe laser transmission at time domain for three indicated signal fields $E_{Sig}$ with fixed LO field power. (b) Measurement of the weak signal field using the resonant EIT-AT splitting technique with $f_{LO}$ = 3.228~GHz and $\delta$ = 75~kHz. The data is taken with the resonant heterodyne technique (gray squares) and EIT-AT splitting in strong field regime (red dots). The red solid line displays the electric field strength calculated by the equation of $E = (30Pg)^{1/2}/d$. The dash line shows the minimum measured electric field of $E_{min}$ = 1.45~$\mu$V/cm.}
		\label{Fig.3}
	\end{figure}

    We then lock the probe laser to the transition $|6S_{1/2}(F = 4)\rangle\to |6P_{3/2}(F^\prime = 5)\rangle$, and the coupling laser to the most sensitivity peak at $m_j$ = 1/2 to measure the response of system to the signal field. The inset of Fig.~\ref{Fig.3} (a) presents the measured transition of the probe laser for three indicated signal fields $E_{Sig}$. As mentioned above, the transmitted EIT signal shows a sinusoidal profile with the frequency $\delta$ = 75~kHz, and the amplitude of sinusoidal curve increases with the signal field power. We measure the amplitude of signal fields as a function of the microwave source power $P$ using the spectrum analyzer. The results are shown as gray squares in Fig.~\ref{Fig.3} (a), the output of the spectrum analyzer linearly increases with the signal source power. To calibrate the signal field values, we measure the Stark shifts of energy at strong field region, in which the $E^2_{tot}$ is proportional to the output power $P$, as expected in Eq.~\ref{Eq.4}. The measured electric field is shown with red dots in Fig.~\ref{Fig.3} (a). The E-field strength at distance of 14~cm from the Rydberg sensor was calculated using 
	\begin{equation}\label{Eq.5}
		E^2=\frac{30P\cdot g}{d^2},
	\end{equation}
    where $E$ is the electric field,  $g$ is a gain factor of the antenna, $d$ is the distance between the horn antenna port and the center of the cesium cell. The calculated result is plotted with a red solid line to calibrate the measurements of Fig.~\ref{Fig.3} (a). The measurements demonstrate a linear dependence on $P$, from which the electric field strength is extracted. The polarizability of $|60D_{5/2}, m_{j} = 1/2\rangle$ state is -4.992~GHz$\cdot$cm$^2$/V$^2$ ~\cite{25}. After the calibration, the minimum detectable value of the signal field is $E_{min}$ = 2.31~$\mu$V/cm, and the linear response dynamic range is over 65~dB.

    For comparison of the performance of the far off-resonant measurement, we do the measurement using the same setup but employing  the resonant EIT-AT spectrum technique~\cite{19}, where the LO field is resonant with the transition $|60D_{5/2}\rangle\to |61P_{3/2}\rangle$ with frequency 3.228~GHz, the measurement is displayed in the Fig.~\ref{Fig.3} (b). The minimum detectable electric field  is $E_{min}$ = 1.45~$\mu$V/cm. More details see our previous work~\cite{26}. From Figs.~\ref{Fig.3} (a) and (b), we can see that the off-resonant AC Stark measurement has a comparable sensitivity with the resonant EIT-AT measurement.

	\begin{figure}[htbp]
		\vspace{-1ex}
		\centering
		\includegraphics[width=0.35\textwidth]{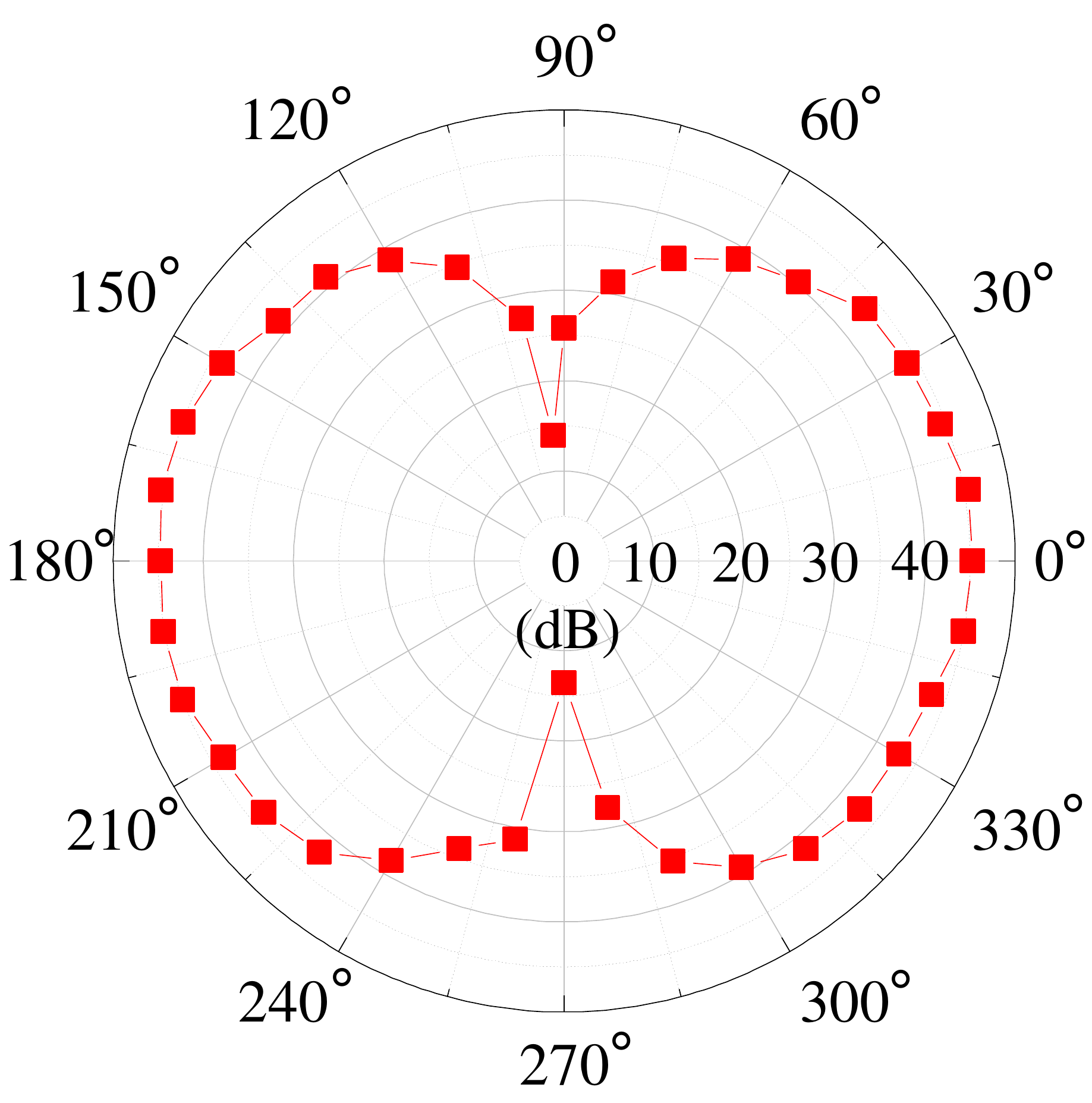}
		\vspace{-1ex}
		\caption{ The measurement of the dependence of output signal of the spectrum analyze on the angle $\theta$ between the LO field and the signal field. The results shows the isolation for two orthogonal signal field is up to 32~dB.}
		\label{Fig.4}
	\end{figure}
	
    To verify the continuous-frequency measurement property of our system, we do a series of measurements like in Fig.~3(a) but with several far off-resonant RF frequencies.  
    The performance of our Rydberg system at different frequency is shown in Tab.~\ref{Tab.1}, where we list the minimum detectable RF electric fields. The detectable field is almost same at the RF frequency range of 2~-~5~GHz. Therefore, our method can achieve the continuously tunable frequency electric field measurement with a high sensitivity. Our measurement of this work presents the RF frequency range below 5~GHz, which is limited by the microwave source we have. In theory, the RF frequency range can be extended to any broad.
	
\begin{table}[htbp]
	\caption{The achieved minimum detectable field at different frequency. }
	\centering
	\begin{center}
		\renewcommand{\arraystretch}{2.5}
    \vspace{0ex}
    \begin{tabular}{|c|c|c|c|c|}
		\hline
		~$f_{LO}$ (GHz)      \quad & \quad $2.18$ \quad\quad & \quad $3.228$ \quad\quad & \quad$4.27$ \quad\quad & \quad $5.18$ \quad~\\
		\hline
		~$E_{min}$($\mu$V/cm) \quad & \quad $2.31$ \quad\quad & \quad $1.45$ \quad\quad & \quad$3.91$ \quad\quad & \quad $8.81$ \quad~\\
		\hline
    \end{tabular}
	\vspace{-3ex}
    \end{center}
    \label{Tab.1}
\end{table}
	
    Finally, We show our system has an inherent polarization selectivity feature as the response of state with different $m_j$ depend on the polarization of incident field ~\cite{4}. Here, we fix the signal field $E_{Sig}$ = 0.73~mV/cm and hold the polarization of the LO field along $y$-axis, while varying the polarization of the signal field by rotating the antenna of the  signal field  around $z$-axis to study the polarization selectivity feature. Fig.~\ref{Fig.4} shows the dependence of output signal from the spectrum analyze on the angle $\theta$ between the LO field and the signal field. We can see that when the signal field and the LO field are co-linear polarization, $\theta$ = 0$^\circ$ or 180$^\circ$, the output signal is 45~dB above the noise. However, when the LO field and signal field are orthogonal, $\theta=90^\circ$ or 270$^\circ$, the output signal is 13~dB above the noise. The isolation for two orthogonal signal field is up to 32~dB.

\section{Conclusion and discussion}
	
	In this paper, we demonstrate the weak electric field measurement with continuous-frequency of 2~GHz-5~GHz using a far off-resonant AC Stark effect in a room-temperature cesium vapor cell. The EIT spectrum is used as an all-optical field probe. The nD Rydberg EIT spectra exhibit $m_j$ dependent AC Stark shifts and splitting at a strong LO field, and the $m_j$ = 1/2 Stark level  exhibits a high sensitivity to the signal field due to its larger polarizability. We achieve a minimum detectable signal field of 2.31~$\mu$V/cm and a linear dynamic range over 65~dB, which is comparable with a resonant microwave-dressed Rydberg heterodyne receiver in the same system. We also show the system is enable to detection of polarization of a RF field. It is worthy to note that the work here provides the new method to measure the weak RF field of the continuous RF frequency. In theory, the AC-Stark-based weak field measurement can achieve continuous operation over a large frequency range and detect weak signals with the high sensitivity, which has a great prospect in the  microwave electric field metrology and atom-based quantum communication.

\section*{Acknowledgment}
	National Key Research and Development Program of China (2017 YFA0304203); National Natural Science Foundation of China (12120101004, 61835007, 62175136); Changjiang Scholars and Innovative Research Team in University of Ministry of Education of China (Grant No. IRT 17R70); Supported by the Fund for Shanxi 1331 Project.

\section*{Data Availability Statement}
The data that support the findings of this study are available from the corresponding author upon reasonable request.

\section*{References}
\bibliography{main}

\end{document}